\documentclass[journal]{IEEEtran}
\usepackage{amsmath,amsfonts,amssymb}
\usepackage{algorithmic}
\usepackage{array}
\usepackage[caption=false,font=normalsize,labelfont=sf,textfont=sf]{subfig}
\usepackage{textcomp}
\usepackage{stfloats}
\usepackage{url}
\usepackage{verbatim}
\usepackage{graphicx}
\usepackage{balance}

\newtheorem{lemm}{Lemma}

\begin{document}

\title{Treating Interference as Noise in Cell-Free Massive MIMO Networks}
\author{Shuaifei~Chen$^{\ast \star}$, Jiayi~Zhang$^{\ast \star}$, Zheng~Chen$^\dag$, and Bo~Ai$^{\star\ddag}$\\
{\small $^\ast$School of Electronic and Information Engineering, Beijing Jiaotong University, Beijing 100044, China.}\\
{\small $^\star$Frontiers Science Center for Smart High-speed Railway System, Beijing Jiaotong University, Beijing 100044, China}\\
{\small $^\dag$Department of Electrical Engineering (ISY), Link{\"o}ping University, SE 58183 Link\"{o}ping, Sweden}\\
{\small $^\ddag$State Key Laboratory of Rail Traffic Control and Safety, Beijing Jiaotong University, Beijing 100044, China.}\\\vspace{-2em}
\thanks{
This work was supported in part by the Fundamental Research Funds for the Central Universities under Grant 2021YJS001, in part by National Key R\&D Program of China under Grant 2020YFB1807201, in part by National Natural Science Foundation of China under Grants 61971027, and 61961130391, in part by Beijing Natural Science Foundation under Grant L202013, in part by Natural Science Foundation of Jiangsu Province, Major Project under Grant BK20212002, in part by the Royal Society Newton Advanced Fellowship under Grant NA191006, in part by Frontiers Science Center for Smart Highspeed Railway System.
The work of Z. Chen was supported in part by the Center for Industrial Information Technology (CENIIT) and the Excellence Center at Link{\"o}ping-Lund in Information Technology (ELLIIT).
}%
}

\maketitle


\begin{abstract}
How to manage the interference introduced by the enormous wireless devices is a crucial issue to address in the prospective sixth-generation (6G) communications.
{The treating interference as noise (TIN) optimality conditions are commonly used for interference management and thus attract significant interest in existing wireless systems.}
Cell-free massive multiple-input multiple-output (CF mMIMO) is a promising technology in 6G that exhibits high system throughput and excellent interference management by exploiting a large number of access points (APs) to serve the users collaboratively.
{In this paper, we take the first step on studying TIN in CF mMIMO systems from a stochastic geometry perspective by investigating the probability that the TIN conditions hold with spatially distributed network nodes.}
We propose a novel analytical framework for TIN in a CF mMIMO system with both Binomial Point Process (BPP) and Poisson Point Process (PPP) approximations.
We derive the probability that the TIN conditions hold in close form using the PPP approximation.
Numerical results validate our derived expressions and illustrate the impact of various system parameters on the probability that the TIN conditions hold.
\end{abstract}
%
%

\section{Introduction}

Densification of the network infrastructure is one of the most essential factors that promotes the substantial development of wireless communications \cite{saad2019vision}.
Especially in the prospective sixth-generation (6G) communications, enormous wireless devices collaborate to offer almost ubiquitous communication service with high network throughput, but meanwhile make interference management become an inevitable challenge, especially in a large-scale network \cite{zhang2019multiple}.
To effectively suppress interference, it is of great importance to analyze the interference behavior of wireless networks.

The treating interference as noise (TIN) optimality conditions are used in wireless communications to determine when the interference is so low that the non-linear transmission schemes are not needed \cite{geng2015optimality}.
These conditions hold if the strength of the desired signal from an intended base station (BS) to an intended user equipment (UE) is greater than or equal to the product of the strengths of the strongest interference that the intended BS creates and of the strongest interference that the intended UE receives.\
The TIN conditions were originally studied for a $K$-user interference channel from an information-theoretic view \cite{geng2015optimality}.
Then the analysis has been extended to the case with device-to-device (D2D) communications \cite{naderializadeh2014itlinq} and cellular networks \cite{bacha2019treating}.
However, there is a research gap here in investigating the probability that the TIN conditions hold, which is referred to as {\it probability of TIN conditions} for simplification in the remainder of this paper, in CF mMIMO using stochastic geometry \cite{haenggi2009interference}.

Cell-free massive multiple-input multiple-output (CF mMIMO) integrates the best parts of cellular mMIMO and ultra-dense network, coordinates a large number of distributed access points (APs) to serve the UEs with almost uniform service quality and excellent interference management \cite{zheng2021uav}, and thus is recognized as a promising paradigm for 6G networks \cite{cellfreebook,chen2021survey,bjornson2019making,zhang2021improving}.
The large-scale deployment of distributed APs brings chances for CF mMIMO to perform collaborative coherent transmission among the APs to suppress the interference, but meanwhile complicates the interference pattern of the system.
A large body of research have investigated the interference management in CF mMIMO on the aspects of performance analysis \cite{chen2018channel,ozdogan2019performance,wang2020uplink}, signal processing \cite{nayebi2017precoding,bjornson2019making,zhang2021local}, and resource allocation \cite{demir2020joint,chen2020structured}.
Especially, \cite{chen2018channel} studied channel hardening and favorable propagation in CF mMIMO networks, which are the biggest virtues of mMIMO systems for interference management, with the tools of stochastic geometry.
To our best of knowledge, TIN has not been considered in the literature of CF mMIMO, which motivates this work.
We summarize the major contributions as follows.
\begin{itemize}
  \item We propose an analysis framework for the probability of TIN conditions with a stochastic geometry approach.
  \item We model the node distributions with both Binomial Point Process (BPP) and Poisson Point Process (PPP).
  \item We derive the probability of TIN conditions in closed form under the PPP approximation, which reveals the system insights in terms of various system parameters via the analytical results.
\end{itemize}

{\emph {Notation}}:
Calligraphic uppercase letters, $\cal A$, denote sets.
$|{\cal A}|$ denotes the cardinality of set $\cal A$.
$b(x,r)$ denotes the $2$-dimensional ball of radius $r$ centered at $x$.
The circularly symmetric complex Gaussian distribution with zero mean and variance $\sigma^2$ is denoted as ${\cal N}_{\mathbb C}\left(0, \sigma^2\right)$.

\section{System Setup}\label{sec:system}

We consider a CF mMIMO system in a finite area $\cal A$ consisting of $L$ single-antenna APs and $K$ single-antenna UEs.
As illustrated in Fig.~\ref{fig:system}, the APs are independently uniformly distributed on the two-dimensional Euclidean plane according to a Binomial Point Process (BPP) ${\bf \Phi}_{\rm ap}$.
Similarly, the locations of the UEs are generated by another independent homogeneous BPP ${\bf \Phi}_{\rm ue}$.
The APs connected via fronthaul connections to a central processing unit (CPU), which is responsible for coordinating and processing the signals of all UEs.
The user-centric CF architecture is adopted \cite{zheng2021impact}, where each UE accesses a subset of APs based on the channel condition.
We let ${\cal M}_k \subset \left\{{ 1,\ldots,L }\right\}$ denote the subset of APs accessed by UE $k$.

To invoke the stationarity of this setup, we consider having a typical UE $0$ at the origin and let the spatially averaged network statistics seen at this typical UE represent the average network performance seen by a randomly located UE.\footnote{In a finite-size network, the statistics seen at network-center users would be different from the network-edge ones. We assume that the network region is large enough with wrap-around effect so that the typical user at the origin can represent a randomly located user in the network.}
Also, we consider a typical AP $0$ located at a distance of $r$ from UE $0$.
{We assume that a UE accesses the APs near than AP $0$, which are referred as its {\it associated APs} such that ${\cal M}_k = \{l: l \in b({\rm UE}_k , r),\ \forall l\}$, $\forall k$, where $b({\rm UE}_k , r)$ is called as the {\it influence region} of UE $k$.
By this, the performance of considering any associated AP is lower bounded by the one of AP $0$.}
For the rest of this paper, we assume that UE $0$ and AP $0$ are included in the considered BPP and area $\cal A$ is a circular region centered at UE $0$ with radius $R$.


\subsection{Channel Model}

We denote by ${h}_{kl} \in {\mathbb C}$ the channel between AP $l$ and UE $k$.
The standard block fading model is adopted, where ${h}_{kl}$ is constant in time-frequency blocks of $\tau_c$ channel uses \cite{bjornson2017massive}.
In each block, the channels are assumed to be subject to Rayleigh fading, i.e.,
\begin{equation}
  {h}_{kl} \sim {\cal N}_{\mathbb C} ({ 0}, {\beta(d_{kl})}),
\end{equation}
where $\beta(d_{kl}) = d_{kl}^{-\alpha}$ is the large-scale fading coefficient that describes the distance-depended pathloss, $d_{kl}$ is the distance between AP $l$ and UE $k$, and $\alpha > 1$ is the pathloss exponent.

To perform coherent transmission, $\tau_p \le \tau_c$ channel uses are exploited for channel estimation such that at most $\tau_p$ orthogonal pilot sequences are available in the system.
Due to the natural channel variations in the time and frequency domain, we have $\tau_p <K$ in most practical scenarios.
In a large-scale CF mMIMO system, it becomes inevitable that some of the UEs might share the same pilot sequence, thus creating interference between each other.

\subsection{Binomial Point Process}

The BPP models the random patterns produced by independently {\it uniformly} distributing a finite number of nodes in a finite service area.
We denote by $d_n$ the random variable representing the distance from an arbitrary reference point $x$ to the $n$-th nearest node.
For a BPP with $N$ randomly scattered nodes in a finite area $\cal W$, the cumulative distribution function (CDF) of $d_n$ is the probability that there are more than $n$ points in the region of $b(x,r)$, as \cite{srinivasa2009distance}
\begin{equation}\label{eq:ccdf_bpp}
  {{F}}_{d_n}(r) = 1 - {I}_{1 - p}(N - n + 1, n), \quad 0 \le r \le R,
\end{equation}
where $p = \frac{|b_d(x,r) \cap {\cal W}|}{|\cal W|}$ and ${I}_{\cdot}(\cdot, \cdot)$ is the normalized incomplete beta function \cite[Eq. (8.392)]{Gradshteyn2007}.
The probability density function (PDF) of the distance function $d_n$ is therefore obtained as
\begin{equation}\label{eq:pdf}
  {{f}}_{d_n}(r) = -\frac{{\rm d}{\bar {F}}_{d_n}(r)}{{\rm d}r} = \frac{{\rm d}p}{{\rm d}r}\frac{{(1-p)^{N-n}p}^{n-1}}{{{B}(N-n+1, n)}},
\end{equation}
where ${{B}(\cdot, \cdot)}$ is the beta function \cite[Eq. (8.384.1)]{Gradshteyn2007}.

\begin{figure}[t!]
\centering
\includegraphics[scale=0.8]{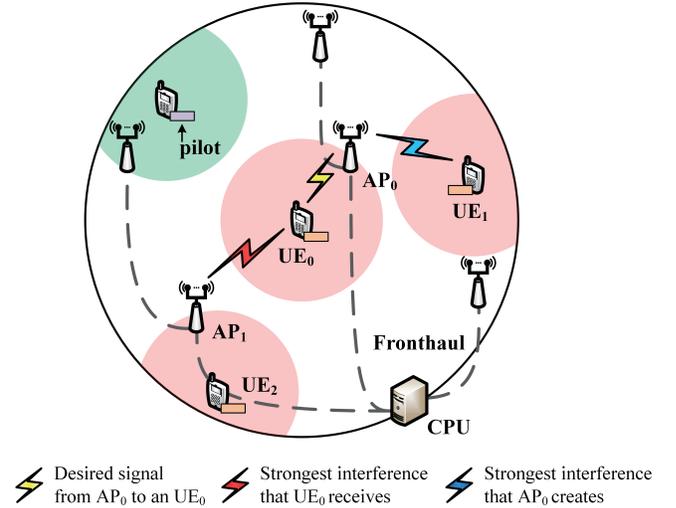}
\caption{TIN in CF mMIMO systems.
\label{fig:system}}
\vspace{0cm}
\end{figure}

\newcounter{longeqn1}
\setcounter{longeqn1}{\value{equation}}
\setcounter{equation}{11}
\begin{figure*}[b]
\normalsize
\hrulefill
\vspace*{4pt}
\begin{equation}\label{eq:derivative_px}
\frac{{\rm d}p_X}{{\rm d}x} =
\begin{cases}
{\frac{2x}{R^2},}&{0<x \le R-r}\\
 {\frac{x}{{\pi Rr\sin {\alpha _{{\rm{ue}}}}}} + \frac{{2x{\alpha _{{\rm{ap}}}}}}{{\pi {R^2}}} - \frac{{{x^2} - {r^2} + {R^2}}}{{2\pi {R^2}r\sin {\alpha _{{\rm{ap}}}}}} - \frac{{{s_\Delta }}}{{\pi {R^2}}}\left( {\frac{1}{{R + r + x}} + \frac{1}{{R - r + x}} + \frac{1}{{r + x - R}} - \frac{1}{{R + r - x}}} \right),}&{ R - r < x \le R + r}
\end{cases}
\end{equation}
\end{figure*}
\setcounter{equation}{\value{longeqn1}}

\subsection{Treating Interference as Noise}

In wireless communications, the TIN optimality conditions characterize the interference relationship between the intended link (from an intended AP to an intended UE), the link between the intended UE and the nearest interfering AP and the link between the intended AP and the nearest interfering UE, as illustrated in Fig~\ref{fig:system}.
{One achieves the whole capacity region to within a constant gap of $\log (3n)$ when the TIN optimality conditions hold, where $n$ is the number of the transceiver pairs \cite{geng2015optimality}.
Mathematically, the TIN optimality conditions in cellular systems can be formulated as \cite{bacha2019treating}}
\begin{equation}\label{eq:tin_condition}
\kappa {\sf{SNR}} ^ \mu \ge \max {\sf{INR}_{{\rm{ap}}}} \cdot \max {\sf{INR}_{{\rm{ue}}}},
\end{equation}
where ${\sf{SNR}}$ denotes the signal-to-noise ratio (SNR) of the intended link, ${\sf{INR}_{{\rm{ap}}}}$ and ${\sf{INR}_{{\rm{ue}}}}$ denote the interference-to-noise ratios (INRs) of the link between the intended AP and the interfering UE and the link between the intended UE and the interfering AP, respectively.
Parameters $\kappa \ge 1$ and $1\le \mu \le 2$ are introduced in \cite{bacha2019treating} for system optimization.

{The TIN conditions in \eqref{eq:tin_condition} does not apply directly in our considered CF mMIMO system since a UE is served by multiple APs instead of one single BS.}
Nevertheless, the ultimate source of interference in multiple antenna systems does not change, that is, the imperfect channel estimates caused by the pilot reuse among the UEs will make coherent transmission less effective, and make it harder to reject interference between UEs assigned with the same pilot.
{With this consideration in mind and letting ${\cal S}_{0}^{\rm ue}$ denote the set of co-pilot UEs of UE $0$ and ${\cal S}_0^{{\rm{ap}}} = \bigcup\nolimits_{k \in {\cal S}_{0}^{\rm ue}} {\cal M}_k / {\cal M}_{0}$, we adapt the TIN conditions in \eqref{eq:tin_condition} for CF mMIMO systems, as}
\begin{equation}\label{eq:tin_condition_cf}
\kappa {\left( {\frac{\rho}{{{\sigma ^2}}}r^{ - \alpha }} \right)^\mu } \ge {\left( {\frac{\rho}{{{\sigma ^2}}}} \right)^2}{\left( {d_{{\rm{ap}}}^{\min }d_{{\rm{ue}}}^{\min }} \right)^{ - \alpha }}
\end{equation}
where $\rho$ is the transmit power of the APs, $\sigma^2$ is the noise power at the UEs, and $d_{{\rm{ap}}}^{\min } = {\min _{k \in {\cal S}_0^{{\rm{ue}}}}}{d_{k0}}$ and $d_{{\rm{ue}}}^{\min } = {\min _{l \in { {\cal S}}_0^{{\rm{ap}}}}}{d_{0l}}$.
Since ${\cal S}_{0}^{\rm ue}$ excludes UE $0$, we refer to ${\cal S}_{0}^{\rm ue}$ as the {\it interfering UEs} set of UE $0$.
Similarly, ${\cal S}_0^{{\rm{ap}}}$ is set of the {\it interfering APs} of UE $0$, in which the APs associate with the co-pilot UEs of UE $0$.
{Note that the interfering UEs interfere with the intended link by deteriorating channel estimation and contaminating the downlink precoder that AP $0$ selects for UE $0$.}
This creates the fundamental difference between the definition of interference in our framework and in existing research with point-to-point links.

\section{Probability that the TIN Conditions Hold}

In this section, for a given intended link between AP $0$ and UE $0$ with link distance $r$, we compute the probability that the TIN conditions in \eqref{eq:tin_condition_cf} hold.
By letting $X = d_{{\rm{ap}}}^{\min }$, $Y = d_{{\rm{ue}}}^{\min }$, and ${g_r} = {{\kappa ^{ - \frac{1}{\alpha }}}{{\left( {\frac{\rho}{{{\sigma ^2}}}} \right)}^{\frac{{2 - \mu }}{\alpha }}}{r^\mu }}$, we represent the probability of TIN conditions ${p}_{\rm tin}$ as
\begin{equation}
 {p}_{\rm tin}
\!=  {\mathbb{P}}\left\{ {X Y \!\ge {g_r}}\right\}{\mathbb{P}}\left\{L_{b({\rm UE}_0, r)} > 0\right\}
\end{equation}
where $L_{b({\rm UE}_0, r)}$ is the number of APs in ${b({\rm UE}_0, r)}$ and
\begin{equation}\label{eq:Lr}
{\mathbb{P}}\left\{L_{b({\rm UE}_0, r)} > 0\right\} = 1 - \left(1 - \frac{r^2}{R^2}\right)^{L}
\end{equation}
is the probability that there exists at least one AP in $b({\rm UE}_0, r)$.
\begin{equation}\label{eq:ptin}
{\mathbb{P}}\left\{ {X Y \!\ge {g_r}}\right\}= 1 -  \int_0^{R+r} \!\!\!{{{f}_X}\left( x \right){{F}_Y}|_r^{\frac{{{g_r}}}{x}}} {\rm{d}}x,
\end{equation}
where ${{f}_X}\left( x \right)$ and ${{F}_Y}\left( y \right)$ as the PDF of $d_{{\rm{ap}}}^{\min }$ and the CDF of $d_{{\rm{ue}}}^{\min }$, respectively.
Note that in \eqref{eq:ptin} the condition probability distribution of $Y | X$ simply becomes $F_Y(y)$ since X and Y are independent.
Also, $X \in [0, R+r]$ and $Y \in [r, R]$.

Although BPP provides a good approximation to practical finite network with finite number of nodes, the analysis of our considered CF mMIMO system is complicated especially when analyzing the exact distribution of $Y$.
Alternatively, we can approximate the distribution of the UEs and the distribution of the APs by equivalent independent homogeneous PPPs with intensity $\lambda_{\rm ue}=\frac{K}{\pi R^2}$ and intensity $\lambda_{\rm ap}=\frac{L}{\pi R^2}$, respectively, when the network size is sufficiently large.

\begin{figure}[t!]
\centering
\includegraphics[scale=0.52]{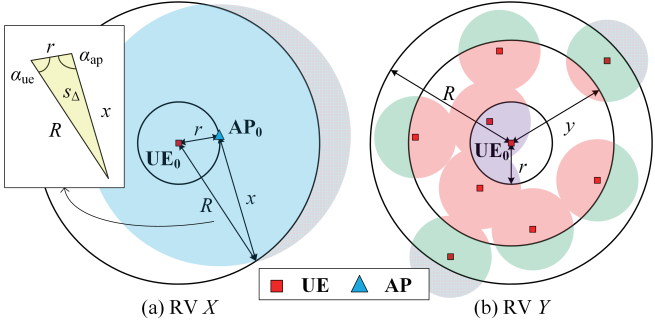}
\caption{Illustrations of (a) $d_{{\rm{ap}}}^{\min }$ (i.e., $X$) and (b) $d_{{\rm{ue}}}^{\min }$ (i.e., $Y$).
\label{fig:XY}}
\vspace{0cm}
\end{figure}

\subsection{PDF of $d_{{\rm{ap}}}^{\min }$}

\subsubsection{BPP-Based Characterization}
Recall that $d_{{\rm{ap}}}^{\min }$ is the distance between the nearest interfering UE and AP $0$.
For a given pilot assignment result, we denote by $|{\cal S}_0^{{\rm{ue}}}| = K'$ the number of the co-pilot UEs in the universal area $\cal A$.
For the reference point AP $0$, the PDF of $d_{{\rm{ap}}}^{\min }$ is obtained with the help of \eqref{eq:pdf} and the fact that $B(n,1) = 1/n$ \cite{Gradshteyn2007}, as
\begin{equation}\label{eq:pdfx}
  {{f}}^{(1)}_X \left( x \right) = \frac{{\rm d}p_X}{{\rm d}x}\frac{(1-p_X)^{K'-1}}{{{B}(K', 1)}} = \frac{{\rm d}p_X}{{\rm d}x}{K'(1-p_X)^{K'-1}},
\end{equation}
where
$
  p_X = \frac{{s}_X}{|\cal A|}
$
is the ratio of the overlapping area ${s}_X = |b({\rm AP}_0, x) \cap {\cal A}|$, illustrated as the blue region in Fig.~\ref{fig:XY}(a), to the area of $|{\cal A}|$.
Since $\cal A$ is finite, $p_X$ is a piecewise function of $x$, as
\begin{equation}\label{eq:px}
  p_X = \begin{cases}
  \frac{x^2}{R^2}, & 0<x \le R-r\\
  \frac{{s}_X}{\pi R^2}, & R-r<x \le R+r
  \end{cases}
\end{equation}
With the help of Heron's formula and cosine rule, the area of the overlapping overage ${s}_X$ is calculated as
\begin{equation}\label{eq:sx}
{s}_X = R^2 \alpha_{\rm ue} + x^2 \alpha_{\rm ap} - 2 {s}_\Delta,
\end{equation}
where $\alpha_{\rm ue} = \arccos \frac{r^2 + R^2 - x^2}{2 r R}$, $\alpha_{\rm ap} = \arccos \frac{r^2 + x^2 - R^2}{2 r x}$, and ${\sf s}_\Delta = \frac{\sqrt{(R+r+x)(R-r+x)(R+r-x)(r+x-R)}}{4}$ (see Fig.~\ref{fig:XY}(a)).
Consequently, the derivative of $p_X$ with respect to $x$ is calculated as \eqref{eq:derivative_px} on the bottom of this page.
\addtocounter{equation}{1}

By substituting \eqref{eq:px}, \eqref{eq:sx}, and $\eqref{eq:derivative_px}$ into \eqref{eq:pdfx}, the PDF of $d_{{\rm{ap}}}^{\min }$ under BPP-based characterization, i.e., ${{f}}^{(1)}_X$, is obtained in closed form.

\subsubsection{Alternative PPP Approximation}
By considering an equivalent PPP $\Phi'_{\rm ue}$ with intensity $\lambda_{\rm ue}$ to approximate the distribution of the UEs, the PDF of $d_{{\rm{ap}}}^{\min }$ can be much simplified.

With random pilot assignment, each UE chooses one pilot sequence uniformly at random from the set of pilot sequences.
Then each sequence is chosen with probability $1/\tau_p$.
Under a given UE intensity $\lambda_{\rm ue}$, the density of co-pilot UEs of UE $0$ is
\begin{equation}
    \lambda_0=1/\tau_p \lambda_{\rm ue}.
    \label{eq:lambda0}
\end{equation}
Thus, the PDF of $d_{{\rm{ap}}}^{\min }$ under PPP approximation is obtained in closed form, as
\begin{equation}\label{eq:pdfx_ppp}
    {{f}}^{(2)}_X \left( x \right) = 2\pi \lambda_0 x \exp(-\pi \lambda_0 x^2).
\end{equation}

\subsection{CDF of $d_{{\rm{ue}}}^{\min }$}

\subsubsection{BPP-Based characterization}
Recall that $d_{{\rm{ue}}}^{\min } = {\min _{l \in { {\cal S}}_0^{{\rm{ap}}}}}{d_{0l}}$ is the distance between the nearest interfering AP and UE $0$.
Since ${\cal S}_0^{{\rm{ap}}} = \bigcup\nolimits_{k \in {\cal S}_{0}^{\rm ue}} {\cal M}_k / {\cal M}_{0}$, an interfering AP should be located within the union influence regions of the co-pilot UEs (denoted by $s_u$ and illustrated as the union circular areas centered at the little red squares in Fig.~\ref{fig:XY}(b)), which intensely baffles the exact analysis of the distribution of $Y$.
To be specific, given the number of the the interfering APs as $|{ {\cal S}}_0^{{\rm{ap}}}| = L'$ in area $\cal A$, with the help of \eqref{eq:ccdf_bpp}, the probability that no interfering AP exists in $b({\rm UE}_0, y)$ is represented as
\begin{equation}\label{eq:}
  {\mathbb{P}}\{L'_{b({\rm UE}_0, y)} = 0 \} = {I}_{1 - p_Y}(L'-L'_{b({\rm UE}_0, y)}, 1),
\end{equation}
where $L'_{b({\rm UE}_0, y)}$ is the number of interfering APs in ${b({\rm UE}_0, y)}$ and
\begin{equation}\label{eq:}
p_Y = \frac{{s_u} \cup b({\rm UE}_0 , y) / b({\rm UE}_0 , r)}{{s_u} \cup b({\rm UE}_0 , R) / b({\rm UE}_0 , r)},
\end{equation}
is illustrated as the ratio of the red region to the union of the red region and green region in Fig.~\ref{fig:XY}(b).
Thus, the CDF of $d_{{\rm{ue}}}^{\min }$ can be written as
\begin{equation}\label{eq:cdfy_bpp}
\begin{aligned}
  {{F}}_{Y}^{(1)}(y) = 1 - \sum\limits_{L'_{b({\rm UE}_0, y)} =0}^{L'-1} {{L'-1 \choose L'_{b({\rm UE}_0, y)}} \left(\frac{r^2}{R^2}\right)^{L'_{b({\rm UE}_0, y)}}}\\
  \times \left(1 - \frac{r^2}{R^2}\right)^{L'- L'_{b({\rm UE}_0, y)} -1}{\mathbb{P}}\{L'_{b({\rm UE}_0, y)} = 0 \}.
\end{aligned}
\end{equation}
The CDF of $d_{{\rm{ue}}}^{\min }$ under BPP-based characterization, i.e., \eqref{eq:cdfy_bpp}, is extremely difficult be computed in closed form, but can be easily computed using Monte Carlo simulations.

\subsubsection{Alternative PPP Approximation}

An AP interferes with UE $0$ if one of its associated UE is sharing the same pilot as UE $0$. When we have many more APs than UEs in a certain area, this implies that the interfering APs will be clustered around the co-pilot UEs.
If we also approximate the distribution of the APs by an equivalent homogeneous PPP $\Phi'_{\rm ap}$ with intensity $\lambda_{\rm ap}$ and the distribution of the co-pilot UEs by an equivalent PPP with intensity $\lambda_0$, then the distribution of interfering APs (of UE 0) in ${ {\cal S}}_0^{{\rm{ap}}}$ follows a Mat\'{e}rn Cluster Process \cite{afshang2017nearest}. The nearest-neighbor distance distribution is given in \cite[Theo. 2]{afshang2017nearest}.

To simplify our analysis, we view the distribution of the interfering APs as a thinned PPP with thinning probability $p_{\rm th}$.
With random pilot assignment, the thinning probability is given by \cite{haenggi2009interference}
\begin{equation}
    p_{\rm th} = 1-\exp(-\lambda_0 \pi  r^2).
\end{equation}

Note that when analyzing the TIN conditions, we only take into account the set of interfering APs ${{\cal S}}_0^{{\rm{ap}}}$ outside the influence region of UE 0. This suggests that $d_{{\rm{ue}}}^{\min }$ is at least $r$.
Then the distribution of $d_{{\rm{ue}}}^{\min }$ can be summarized as
\begin{equation}
f_Y(y) = \begin{cases}
0, &\text{if $y<r$}\\
c \cdot 2\pi p_{\rm th}\lambda_{\rm ap} y \exp(-\pi p_{\rm th}\lambda_{\rm ap} y^2),  &{\rm otherwise}
\end{cases}
\end{equation}
where $c = \exp \left( {  \pi p_{\rm th}\lambda _{\rm ap} r^2} \right)$ is a normalization factor such that the integration of $f_Y(y)$ over $[0,\infty]$ is $1$.
With the help of \cite[Eq. (3.321.4)]{Gradshteyn2007}, the CDF of $d_{{\rm{ue}}}^{\min }$ can be calculated in closed form, as
\begin{equation}\label{eq:cdfy_ppp}
{F_Y^{(2)}}(y) = \begin{cases}
 {0,}&{{\rm if}\ y<r}\\
{1 - \exp \left( {\pi {p_{{\rm{th}}}}{\lambda _{{\rm{ap}}}}\left( {{r^2} - {y^2}} \right)} \right)
,}&{\rm otherwise}
\end{cases}
\end{equation}

\begin{figure}[t!]
\centering
\includegraphics[scale=0.63]{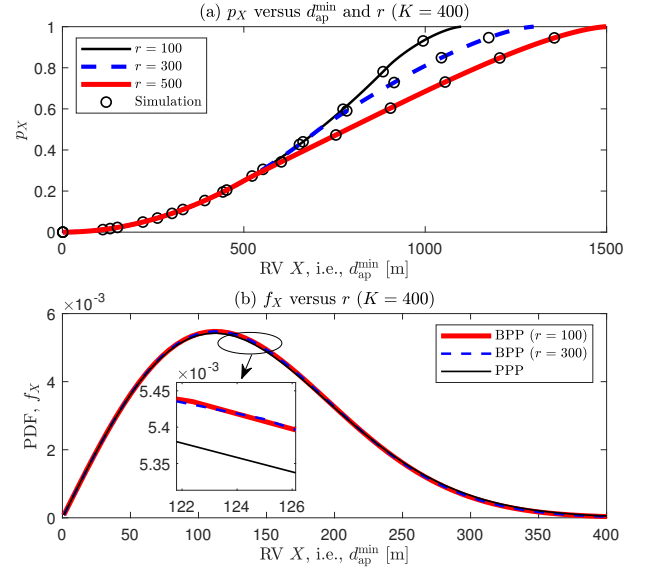}
\caption{PDF $f_X$ and $p_X$ versus $X$ (i.e., $d_{\rm ap}^{\min}$) with different $r$s.
\label{fig:x}}
\vspace{0cm}
\end{figure}

\begin{figure}[t!]
\centering
\includegraphics[scale=0.63]{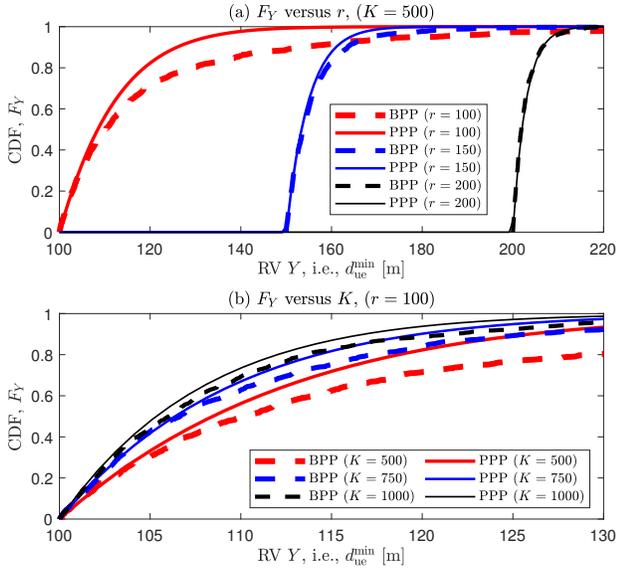}
\caption{CDF $F_Y$ versus $Y$ (i.e., $d_{\rm ue}^{\min}$) with different $r$ and $K$.
\label{fig:y}}
\vspace{0cm}
\end{figure}

\begin{figure}[t!]
\centering
\includegraphics[scale=0.63]{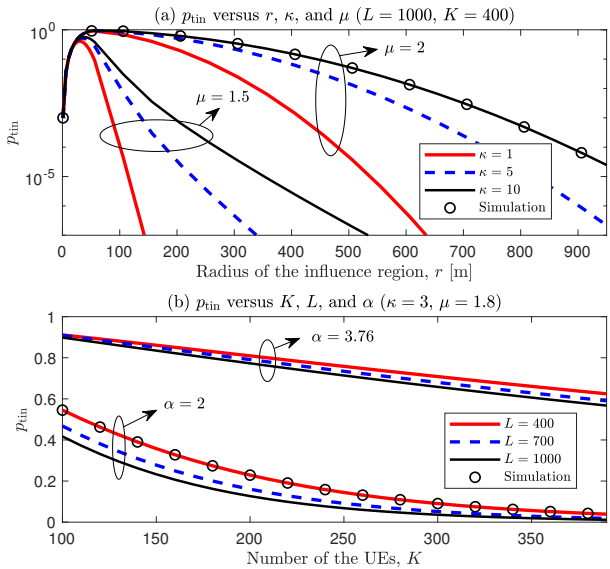}
\caption{Probability of TIN conditions $p_{\rm tin}$.
\label{fig:ptin}}
\vspace{0cm}
\end{figure}

\newcounter{longeqn2}
\setcounter{longeqn2}{\value{equation}}
\setcounter{equation}{23}
\begin{figure*}[b]
\normalsize
\hrulefill
\vspace*{4pt}
\begin{equation}\label{eq:ptin2}
{\mathbb{P}}\left\{ {X Y \!\ge {g_r}}\right\} = \exp \left( { - \pi {\lambda _0}{{\left( {\frac{{{g_r}}}{r}} \right)}^2}} \right) \!+ \!\exp \left( {\pi {p_{{\rm{in}}}}{\lambda _{{\rm{ap}}}}{r^2}} \right)\pi {\lambda _0}\left(\! {1 - \frac{{g_r^2}}{{{r^2}}}G_{11;10;01}^{10;01;10}\left( {\begin{array}{*{20}{c}}
{ - 1}\\
0
\end{array}\left| {\begin{array}{*{20}{c}}
1\\
{}
\end{array}\left| {\left. {\begin{array}{*{20}{c}}
{}\\
0
\end{array}} \right|\frac{{{r^2}}}{{g_r^2\pi {\lambda _0}}},\pi {p_{{\rm{in}}}}{\lambda _{{\rm{ap}}}}{r^2}} \right.} \right.} \right)} \right)
\end{equation}
\end{figure*}
\setcounter{equation}{\value{longeqn2}}

\subsection{Closed-Form Expression of $p_{\rm tin}$}

By substituting \eqref{eq:pdfx_ppp} and \eqref{eq:cdfy_ppp} into \eqref{eq:ptin}, we rewrote \eqref{eq:ptin} as
\begin{align}\label{eq:ptin1}
\notag &{\mathbb{P}}\left\{ {X Y \!\ge {g_r}}\right\} \\
\notag =& 1 \!-\! 2\pi {\lambda _0}\!\left( \!{\int_0^{\frac{g_r}{r}} {\!\!\!\!\!x\exp \left( { - \pi {\lambda _0}{x^2}} \right){\rm d}x}  \!-\! \exp \left( {\pi {p_{{\rm{in}}}}{\lambda _{{\rm{ap}}}}{r^2}} \right)} \right.\\
 &\times \left. {\int_0^{\frac{g_r}{r}} {x\exp \left( { - \pi {\lambda _0}{x^2} - \frac{{\pi {p_{{\rm{in}}}}{\lambda _{{\rm{ap}}}}g_r^2}}{{{x^2}}}} \right){\rm d}x} } \right)\\
\notag =& 1 - 2\pi {\lambda _0}\left( {{{\cal I}_1} - \exp \left( {\pi {p_{{\rm{in}}}}{\lambda _{{\rm{ap}}}}{r^2}} \right) {{\cal I}_2}} \right).
\end{align}
Note that the upper limit of the integral in \eqref{eq:ptin1} is $\frac{g_r}{r}$, which differs from the one of the integral in \eqref{eq:ptin}, i.e., $R+r$.
${\cal I}_1$ is calculated with the help of \cite[Eq. (3.321.4)]{Gradshteyn2007}, as
\begin{equation}\label{eq:I1}
{{\cal I}_1} = \frac{1}{{2\pi {\lambda _0}}}\left( {1 - \exp \left( { - \pi {\lambda _0}{{\left( {R + r} \right)}^2}} \right)} \right),
\end{equation}
and ${\cal I}_2$ is calculated in the following lemma.
\begin{lemm}
${\cal I}_2$ can be evaluated in closed form as
\begin{align}\label{eq:I2}
&{{\cal I}_2} = \frac{1}{2} \left(1 \!-\! \int_{\frac{{g_r^2}}{{{r^2}}}}^\infty  {\exp \left( { - \pi {\lambda _0}t - \frac{{\pi {p_{{\rm{in}}}}{\lambda _{{\rm{ap}}}}g_r^2}}{t}} \right)dt}\right) \\
\notag &= \frac{1}{2}\left( {1 - \frac{{g_r^2}}{{{r^2}}}G_{11;10;01}^{10;01;10} \!\left( \!\!\begin{array}{c|c|c|}
-1& 1 & \\
0 & & 0 \\
\end{array} \ {\frac{{{r^2}}}{{g_r^2\pi {\lambda _0}}},\pi {p_{{\rm{in}}}}{\lambda _{{\rm{ap}}}}{r^2}} \right) } \!\!\right)
\end{align}
where $G_{-}^{-}\left( \cdot \right)$ denotes the extended generalized bivariate Meijer $G$-function (EGBMGF) \cite{Shah1973On}.
\end{lemm}
\begin{IEEEproof}
It follows the similar approach as in \cite[Appe.]{chen2018performance} and is omitted due to the space limitation.
\end{IEEEproof}

By substituting \eqref{eq:I1} and \eqref{eq:I2} into \eqref{eq:ptin1}, ${\mathbb{P}}\left\{ {X Y \!\ge {g_r}}\right\}$ is obtained in closed form as \eqref{eq:ptin2} on the bottom of next page.
With \eqref{eq:Lr} and \eqref{eq:ptin2}, we can straightforwardly obtain the closed-form expression of $p_{\rm tin}$.

\section{Numerical Results}\label{sec:results}

In this section, we quantitatively validate our derived theoretical results and therefore evaluate the probability of TIN conditions in our considered CF mMIMO systems.
We consider a finite circular coverage area with radius $R =1$ km in which the APs and UEs are scattered uniformly at random.
The well-known 3GPP Urban Microcell model to compute the large-scale fading coefficients.
For the fair transformation between the BPP and PPP setups, we let $K' = K/ {\tau_p}$ and such that $\lambda_0 \pi R^2 = K'$.
Unless otherwise specified, the following default parameters: $\tau_p = 10$, $L = 1000$, $\alpha = 3.76$, $\rho=1$ W, and $\sigma^2 = -94$ dBm.

We first evaluate our derived analytical results for $d_{{\rm{ap}}}^{\min }$ with $K=400$.
Fig.~\ref{fig:x}(a) validates our derived closed-form expression for $p_X$ in \eqref{eq:sx} ($p_X$ is straightforwardly obtained from $s_X$), where the ``Simulation" results are generated by performing Monte Carlo simulations.
Fig.~\ref{fig:x}(b) shows that the PPP can excellently approximate BPP, where the ``BPP" and ``PPP" results are generated by \eqref{eq:pdfx} and \eqref{eq:pdfx_ppp}, respectively.
It is worth noting that $r$ barely effects $f_X^{(1)}$ in Fig.~\ref{fig:x}(b) due to the large-scale network deployment ($R \gg r$).

Fig.~\ref{fig:y} demonstrates the tightness of the PPP approximation to BPP when dealing with $d_{{\rm{ue}}}^{\min }$.
The first observation is that the gap from the PPP approximation to BPP decreases as either $K$ or $r$ increases.
This comes from the fact that with a small influence region radius $r$, the interfering APs are highly coupled with the co-pilot UEs, causing their behavior is similar to a Mat\'{e}rn Cluster Process.
When $r$ goes large, the impact of point process fades away, the distribution of $d_{{\rm{ue}}}^{\min }$ becomes deterministic, which makes the CDF curves steep.
From $K$'s point of view, a larger intensity $\lambda_{\rm ue}$ promotes the randomness of the UEs in the system, which tightens the gap between the PPP approximation and the BPP.

Since Fig.~\ref{fig:x} and Fig.~\ref{fig:y} have validated the tight approximation of the PPP to the BPP, we can safely use $f_X^{(2)}$ and $F_Y^{(2)}$ to compute the probability of TIN conditions $p_{\rm tin}$ in our considered large-scale CF mMIMO systems.
In Fig.~\ref{fig:ptin}, we first validate our derived closed-form expression of $p_{\rm tin}$ by computing $p_{\rm tin}$ with \eqref{eq:ptin} and \eqref{eq:ptin2}, respectively.
Then, we analyze the performance of $p_{\rm tin}$ with respect to several system parameters, i.e., $K$, $L$, $r$, $\kappa$, $\mu$, and $\alpha$.
In Fig.~\ref{fig:ptin}(a), the first observation is that $p_{\rm tin}$ is a concave function with respect to $r$, where significant deteriorations in $p_{\rm tin}$ appear when $r$ approaches either $0$ or $R$.
The reason is that no AP will locate at the influence region $b({\rm UE}_0 , r)$ when $r$ is extremely small, which forces ${\mathbb{P}}\left\{L_{b({\rm UE}_0, r)} > 0\right\},\ p_{\rm tin} \to 0$; on the opposite, a too large $r$ will force ${\sf{SNR}} \approx {\sf INR}_{\rm ue}$ while $\max {\sf INR}_{\rm ap} \gg {\sf{SNR}}$, causing $p_{\rm tin } \to 0$ (note that $1\le \mu \le 2$).
Moreover, as exhibited in \eqref{eq:tin_condition} and \eqref{eq:tin_condition_cf}, Fig.~\ref{fig:ptin}(a) shows that $\kappa$ and $\mu$ can promote $p_{\rm tin}$ with the latter having a much more significant impact.
In Fig.~\ref{fig:ptin}(b), we notice that $p_{\rm tin}$ decreases as $K$ increases since a larger UE intensity $\lambda_{\rm ue}$ raises both $\max {\sf INR}_{\rm ap}$ and $\max {\sf INR}_{\rm ue}$.
The interesting observation is that a larger AP intensity $\lambda_{\rm ap}$ also reduces $p_{\rm tin}$, which seems contrary to the conclusions in the prior studies on CF mMIMO that a larger AP intensity will promote the system performance.
This comes from the fact that $p_{\rm tin}$ only characterize the strength of one AP at the edge of $b({\rm UE}_0 , r)$, not the sum strength of all AP within $b({\rm UE}_0 , r)$.
Thus, a larger AP intensity $\lambda_{\rm ap}$ will only increase $\max {\sf INR}_{\rm ue}$ with no positive effect on ${\sf{SNR}}$, which reduces $p_{\rm tin}$.
For the similar reason as $L$, pathloss exponent $\alpha$ promotes $p_{\rm tin}$ instead of reducing it since $\alpha$ has more significant impact on the product of $\max {\sf INR}_{\rm ap}$ and $\max {\sf INR}_{\rm ue}$ than ${\sf{SNR}}$, which suggests that $\max {\sf INR}_{\rm ap}\cdot \max {\sf INR}_{\rm ue}$ decreases faster that ${\sf{SNR}}$ when $\alpha$ goes large.

\section{Conclusion}\label{sec:conclusion}

We investigated the probability of TIN conditions in a large-scale CF mMIMO system with randomly distributed APs and UEs.
We first proposed an analysis framework for the probability of TIN conditions by considering BPP to model the node distributions.
To perform a tractable analysis, we further introduced an alternative PPP approximation, which characterizes the distribution of contact distance of the nearest interfering AP and nearest interfering UE in closed form.
Consequently, we derived the closed-form expression for the probability of TIN conditions.
Numerical results validated our derived expressions and indicated the parameter interval for tight approximation.
Also, the performance of the probability of TIN conditions under various system parameters is analyzed.
Specifically, the probability of TIN optimality conditions increases as the TIN parameters $\kappa$ and $\mu$ and channel parameter $\alpha$ increase, and this probability decreases as the UE number $K$ and the AP number $L$ increase.
By appropriately adjusting the influence region radius $r$, an optimal probability of TIN conditions can be achieved.
The results are beneficial for the performance analysis of TIN and the design of the TIN-based interference management schemes for practical CF mMIMO systems, which is left for our future work.

\bibliographystyle{IEEEtran}
\bibliography{IEEEabrv,Ref}
\end{document}